\documentclass[conference]{IEEEtran}
\IEEEoverridecommandlockouts
\usepackage{cite}
\usepackage{amsmath,amssymb,amsfonts}
\usepackage{graphicx}
\usepackage{textcomp}
\usepackage{xcolor}
\usepackage{float} 
\usepackage{booktabs}
\usepackage{url}
\usepackage{multirow}
\usepackage[ruled,vlined,linesnumbered]{algorithm2e}
\usepackage[flushleft]{threeparttable}
\DeclareMathSizes{10}{9}{7}{5}
\def\BibTeX{{\rm B\kern-.05em{\sc i\kern-.025em b}\kern-.08em
    T\kern-.1667em\lower.7ex\hbox{E}\kern-.125emX}}
\begin{document}
\raggedbottom
\title{Closed-Loop Decision-Focused Learning for User-Aware Cloud Orchestration under
Uncertainty
}

\author{\IEEEauthorblockN{Dongbin~Jiao\IEEEauthorrefmark{1}, Xubo~Zhang\IEEEauthorrefmark{1}, Huakang~Lin\IEEEauthorrefmark{2}, Ke~Shang\IEEEauthorrefmark{3}, and Shi~Yan\IEEEauthorrefmark{1}}
\IEEEauthorblockA{\IEEEauthorrefmark{1}School of Information Science and Engineering, Lanzhou University, Lanzhou, 730000, P. R. China}
\IEEEauthorblockA{\IEEEauthorrefmark{2}Division of Computer Science and Engineering, College of Engineering,
Louisiana State University, \\ Baton Rouge, LA 70803, USA}
\IEEEauthorblockA{\IEEEauthorrefmark{3}School of Artificial Intelligence, Shenzhen University,
Shenzhen, 518060, P. R. China}
Email: \{jiaodb, zhxubo2025\}@lzu.edu.cn,  hlin29@lsu.edu, kshang@foxmail.com, yanshi@lzu.edu.cn}

\maketitle

\begin{abstract}
Time-varying cloud workloads often cause resource under-utilization during off-peak periods and resource contention during peak periods. Existing prediction-then-optimization (PTO) frameworks suffer from two-stage decoupling, hindering the balance among violation rate, user satisfaction, and resource utilization. We formulate heterogeneous job scheduling as a multi-objective combinatorial optimization problem (MOCOP) under uncertain constraints and propose a closed-loop decision-focused learning (CL-DFL) framework for cloud orchestration. CL-DFL integrates a Multivariate Time-series Graph Neural Network (MTGNN)-based spatio-temporal predictor with a zeroth-order decision-focused learning (DFL) mechanism based on the tree-structured Parzen estimator (TPE). This integration establishes an end-to-end (E2E) feedback pathway between resource perception and scheduling decisions.
Furthermore, we develop the GNeuro-PLS strategy by incorporating group relative policy optimization (GRPO) into  cooperative local search to improve robustness under heterogeneous workloads. Extensive experiments on four real-world datasets demonstrate that CL-DFL achieves superior trade-offs among violation rate, user satisfaction, and resource utilization. It  effectively controls overload risks under regular workloads and maintains resilience under highly saturated scenarios compared with state-of-the-art baselines.

\end{abstract}


\section{Introduction}
With the rapid advancement of artificial intelligence (AI) and cloud computing, cloud data centers have become essential infrastructures for supporting large-scale intelligent applications and digital services. Major cloud providers, including Amazon Web Services (AWS), Microsoft Azure, and Google Cloud, offer elastic resource provisioning and on-demand computing services to support diverse workloads \cite{lee2022dynamic}. However, as cloud infrastructures continue to scale, achieving high resource utilization while ensuring service reliability and user quality-of-service (QoS) remains a critical challenge \cite{mesbahi2018reliability,gawali2018task}.


Modern cloud workloads exhibit significant temporal dynamics and resource heterogeneity \cite{zhao2026survey}. The mismatch between fluctuating workload demands and available computing resources often results in resource under-utilization during off-peak periods and resource contention during peak periods. To improve resource efficiency, pre-collected job scheduling has been introduced \cite{dong2021predictive}, where users provide execution flexibility in exchange for lower costs. Unlike latency-sensitive online jobs, pre-collected jobs enable providers to exploit future resource availability for efficient orchestration.
Cloud orchestration requires system-level coordination among workload perception, resource forecasting, scheduling decisions, and service objectives. Therefore, effective cloud orchestration should establish a closed-loop interaction between prediction and optimization rather than relying on isolated prediction-then-optimization (PTO) pipelines.

However, efficient orchestration of heterogeneous cloud jobs remains challenging \cite{yukun2024computing}. Existing methods mainly rely on cluster-level resource prediction, ignoring server-level capacity distributions and leading to inaccurate placement decisions. Moreover, current schedulers often optimize system objectives (e.g., utilization and cost) while overlooking user latency requirements, resulting in QoS degradation and potential service level agreement (SLA) violations. The decoupling between prediction and scheduling also prevents decision feedback from correcting forecasting errors under dynamic workloads.
Furthermore, real-world cloud workloads are inherently heterogeneous in both resource demand and temporal flexibility \cite{he2022online}. Large-scale jobs increase resource contention, while small-scale jobs exploit fragmented resources. Meanwhile, flexible execution windows introduce scheduling flexibility but create trade-offs among capacity violations, user satisfaction, and resource utilization. These challenges motivate the design of a fine-grained adaptive orchestration framework that jointly considers server-level availability, workload heterogeneity, and multi-objective scheduling.

To address these challenges, we propose a closed-loop decision-focused learning (CL-DFL) framework for heterogeneous cloud orchestration. Unlike conventional PTO approaches, such as controlling under uncertain constraints (CUC)~\cite{dong2021predictive}, CL-DFL establishes a feedback-driven interaction between resource forecasting and scheduling decisions. Specifically, we formulate heterogeneous job orchestration as a multi-objective combinatorial optimization problem (MOCOP) with server-level capacity constraints. A spatio-temporal capacity prediction mechanism based on multivariate time-series graph neural network (MTGNN) \cite{wu2020connecting} is adopted to capture correlations among physical servers. Moreover, a zeroth-order decision-focused learning strategy is introduced to adapt forecasting model based on scheduling feedback. Based on the predicted capacity and feedback information, we further design a GNeuro-PLS cooperative evolutionary scheduling strategy that leverages group relative policy optimization (GRPO) \cite{shao2024deepseekmath} to efficiently search Pareto-optimal solutions in large-scale heterogeneous cloud environments. The main contributions of this paper are summarized as follows: (i) We formulate a heterogeneous cloud orchestration model with server-level capacity constraints and a maximum latency tolerance mechanism, establishing a multi-objective optimization (MOO) problem that jointly considers violation rate, user satisfaction, and resource utilization. (ii) We employ MTGNN with adaptive graph learning for capacity prediction to capture inter-server correlations and provide fine-grained resource availability estimation. (iii) We propose a closed-loop DFL framework that incorporates scheduling feedback into prediction adaptation, overcoming the limitations of conventional decoupled PTO paradigms. (iv) We design the GNeuro-PLS strategy to efficiently explore Pareto trade-offs among conflicting objectives in large-scale heterogeneous cloud orchestration. Extensive experiments on real-world cloud workload traces validate the effectiveness of the proposed framework.

\section{Related Work}

\textbf{Cloud Resource Capacity Forecasting}. 
Accurate cloud resource forecasting is essential for efficient scheduling and resource optimization. Early studies mainly relied on statistical approaches  \cite{liu2016quantitative} or deep learning models \cite{christofidi2023machine}, which captured temporal workload patterns but ignored spatial dependencies among distributed servers \cite{feng2024application}. Recently, graph neural networks (GNNs) have been introduced to model such dependencies~\cite{huang2026dyneformer,luo2026causality}. However, most existing methods focus on cluster-level aggregation and lack fine-grained server-level capacity modeling \cite{liEvoGWP2024}. In contrast, MTGNN \cite{wu2020connecting} leverages adaptive graph learning to capture dynamic inter-server correlations without predefined topologies, making it suitable for heterogeneous server-level resource forecasting.

\textbf{Cloud Resource Scheduling and MOO}. 
Pre-collected task scheduling can be formulated as a combinatorial optimization problem that couples resource prediction and scheduling decisions \cite{dong2021predictive}.
Existing PTO-based methods first estimate future resource states and then perform scheduling. 
Although stochastic programming and robust optimization handle uncertainty theoretically, their high computational complexity limits scalability in dynamic cloud environments \cite{shao2022learning,ruparel2025carbon}. To address the increasing need for balancing multiple conflicting objectives, MOO techniques, including evolutionary algorithms (e.g., NSGA-II \cite{deb2002fast}) and heuristic searches, have been explored to balance conflicting objectives. 
However, these approaches often suffer from slow convergence and poor scalability in large-scale scheduling scenarios
\cite{cui2023multi,fang2025energy}. Moreover, predefined weights or relaxation strategies may fail to capture diverse SLA requirements and resource-efficiency trade-offs in heterogeneous clouds.

\textbf{Decision-Focused Learning for Resource Scheduling}.
The decoupling between prediction and optimization limits intelligent resource management, as prediction models in conventional PTO frameworks cannot leverage downstream decision feedback. As a result, improved 
prediction accuracy does not always lead to better scheduling performance under dynamic uncertainties. Learning-to-optimize (L2O) approaches integrate learning models with optimization procedures to improve heuristic search and decision quality \cite{lin2022pareto,zhang2025neuropls,yang2025learning}. Reinforcement learning (RL)-based scheduling methods have also been explored for sequential decision-making but often suffer from training instability, slow convergence, and limited generalization in large-scale constrained optimization \cite{huang2023deep}. Recently, decision-focused learning (DFL) has emerged as a promising paradigm that directly optimizes prediction models according to downstream decision performance \cite{mandi2024decision,wen20263d}. However, applying DFL to cloud orchestration remains challenging due to non-convex optimization, non-differentiable objectives, and complex resource constraints, motivating closed-loop frameworks that jointly adapt prediction and scheduling.


\section{System Model and Problem Formulation}
\label{ch:model}
To overcome the limitations of existing CUC-based approaches, this section presents a heterogeneous cloud model, fine-grained capacity prediction mechanism, and MOO problem formulation for pre-collected tasks scheduling.

\subsection{Cloud Heterogeneous Job Request  Representation}
Cloud task scheduling is typically performed over discrete time intervals. Let $\mathcal{S} = \{s_j \mid 1 \le j \le S\}$ denote a set of $S$ physical servers, and $\mathcal{T} = \{1, 2, \dots, T\}$ represent the scheduling horizon consisting of $T$ discrete time steps. Given a set of pending task requests $\mathcal{B} = \{b_i \mid 1 \le i \le N\}$, where $N$ denotes the number of tasks awaiting scheduling, each task request $b_i$ is characterized by a four-tuple $(c_i, d_i, e_i, l_i)$. Specifically, $c_i$ denotes the computing resource demand of task $i$, covering workloads ranging from lightweight microservices to resource-intensive deep learning jobs, $d_i$ represents the execution duration, capturing the temporal diversity of short-lived bursty tasks and long-running batch jobs, $e_i$ and $l_i$ denote the earliest and latest start times specified by the user, respectively. The temporal flexibility of task $i$ is quantified by the scheduling window $\Delta T_i = l_i - e_i$.

\subsection{Spatio-Temporal Coupled Available Capacity Prediction}


Efficient scheduling requires accurate server-level capacity prediction. Existing aggregated datacenter-level forecasting ignores inter-server resource distributions, leading to inaccurate decisions since sufficient global capacity may not ensure individual server feasibility for heterogeneous tasks.

To address this issue, we model the available capacity prediction problem as a spatio-temporal forecasting task. Let $\mathbf{A}_{\text{past}} \in \mathbb{R}^{S \times T_{\text{in}}}$ denote the historical available capacity matrix,  where  $a_{j, t}^{\text{past}}$ represents the true remaining capacity of server $s_j$ at time slot $t$. The forecasting model aims to learn a mapping function $f_{\theta}(\cdot)$ that predicts the future capacity matrix: 
\begin{equation}
    \hat{\mathbf{A}} = f_{\theta}(\mathbf{A}_{\text{past}}; \mathcal{G}),
\end{equation}
where $\hat{\mathbf{A}} \in \mathbb{R}^{S \times T}$ denotes the predicted available capacity of all servers over the future scheduling horizon, and $\mathcal{G}$ represents the server correlation matrix capturing spatial dependencies among nodes. The predicted capacity matrix provides fine-grained resource availability for scheduling and constrains heterogeneous task placement.

\subsection{Multi-Objective Task Scheduling Formulation}
Given the predicted capacity matrix $\hat{\mathbf{A}}$, the scheduler determines
the execution server and start time for each pre-collected task request in $\mathcal{B}$. We define a
binary decision variable $x_{i,j,t} \in \{0, 1\}$, where $x_{i,j,t}=1$ indicates that task $i$ is assigned on server $j$ at time slot $t$, and $0$ otherwise. The actual scheduled start time of task $i$ is calculated as
\begin{equation}
    t^s_i = \sum_{j=1}^S \sum_{t=1}^T t \cdot x_{i,j,t}.
\end{equation}

To jointly optimize violation rate, user satisfaction, and resource utilization, we formulate the scheduling problem as an MOO problem.

\subsubsection{Objective Functions}

\emph{Violation Rate} ($f_1$):  This objective measures the mismatch between scheduling decision and the actual available capacity $\mathbf{A}$. It is defined as the ratio of overloaded server-time pairs:
\begin{equation}
    f_1 = \frac{1}{S \times T} \sum_{j=1}^S \sum_{t=1}^T \mathbb{I} \left( \sum_{i=1}^N \sum_{t'=t-d_i+1}^t c_i x_{i,j,t'} > a_{j,t} \right),
\end{equation}
where $\mathbb{I}(\cdot)$ is the indicator function and $a_{j,t}$ is the actual available capacity of server $j$ at time slot $t$.

\emph{User Satisfaction} ($f_2$): User satisfaction is determined by the task start time relative to its requested latest start time $l_i$. A task achieves full satisfaction if it starts before $l_i$, while the satisfaction score decreases linearly with the delay duration within the maximum tolerable delay $t_{\text{delay}}$. Unscheduled tasks are assigned a satisfaction score of zero. The overall user satisfaction is defined as 
\begin{equation}
    f_2 = \frac{1}{N} \sum_{i=1}^{N} \left( \sum_{j=1}^S \sum_{t=1}^T x_{i,j,t} \right) \max \left( 0, 1 - \frac{\max(0, t^s_i - l_i)}{t_{\text{delay}} + \epsilon} \right),
\end{equation}
where $N$ is the total number of pre-collected task requests and $\epsilon$ is a small constant to avoid division by zero. The normalization by $N$ ensures that unscheduled tasks directly reduce the overall satisfaction score.


\emph{Resource Utilization} ($f_3$): The resource utilization objective evaluates the effective utilization of available capacity:
\begin{equation}
    f_3 = \frac{\sum_{j=1}^S \sum_{t=1}^T \min \left( u_{j,t}, a_{j,t} \right)}{\sum_{j=1}^S \sum_{t=1}^T a_{j,t} + \epsilon},
\end{equation}
where $u_{j,t} = \sum_{i=1}^{N} c_i \sum_{t' = t - d_i + 1}^{t} x_{i,j,t'}$ represents the total workload assigned to server $j$ at time slot $t$.

\subsubsection{Multi-Objective Scheduling Problem} \label{sec-formulation}
We formulate the task scheduling problem as an MOCOP under uncertain constraints, where $\Omega$ is the discrete feasible solution space and $X \in \Omega$ is a feasible scheduling scheme.
\begin{subequations}
\label{eq:mcop_formulation}
\begin{align}
    \min \quad & \mathbf{F} = [f_1(\mathbf{X}), 1 - f_2(\mathbf{X}), 1-f_3(\mathbf{X})]^T \label{eq:moo_obj} \\
    \text{s.t.} \quad & \sum_{j=1}^S \sum_{t=1}^T x_{i,j,t} \le 1, \quad \forall i \in \{1, \dots, N\}, \label{eq:con_unique} \\
     & x_{i,j,t} = 0, \quad \forall t \notin (e_i, l_i + t_{\text{delay}}), \label{eq:con_time_window} \\
    & u_{j,t} \le \hat{a}_{j,t}, \quad \forall j \in \{1, \dots, S\}, t \in \{1, \dots, T\}, \label{eq:con_capacity} \\
    & x_{i,j,t} \in \{0, 1\}, \quad \forall i, j, t. \label{eq:con_binary}
\end{align}
\end{subequations}
Here, $\mathbf{X}=\{ x_{i,j,t} \mid i=1,\dots,N,\; j=1,\dots,S,\; t=1,\dots,T \}$. Constraints \eqref{eq:con_unique} guarantee that each task is scheduled at most once within its lifecycle, avoiding resource  contention and duplicate execution. Constraint \eqref{eq:con_time_window}  restricts tasks within the earliest start time and the allowed maximum latency tolerance window. Constraint \eqref{eq:con_capacity} incorporates the predicted capacity boundaries $\hat{a}_{j,t}$ obtained from the forecasting module, ensuring that scheduling decisions are generated based on fine-grained server-level resource availability.


\section{Collaborative Optimization Framework Based on Decision-Focused Learning}
\label{sec:dfl_collaborative_framework}
To overcome the limitations of existing PTO approaches, we propose CL-DFL, a closed-loop DFL framework for heterogeneous cloud orchestration. CL-DFL integrates resource perception, decision optimization, and feedback adaptation into an end-to-end (E2E) closed-loop pipeline. The framework consists of three key components: an MTGNN-based spatio-temporal perception module, a GNeuro-PLS cooperative scheduling strategy, and a zeroth-order DFL mechanism based on the tree-structured Parzen estimator (TPE) \cite{watanabe2023c}.

\subsection{CL-DFL Collaborative Optimization Framework}
\label{subsec:icuc_overall_design}
\begin{figure*}[ht]
    \centering        \includegraphics[width=0.9\linewidth]{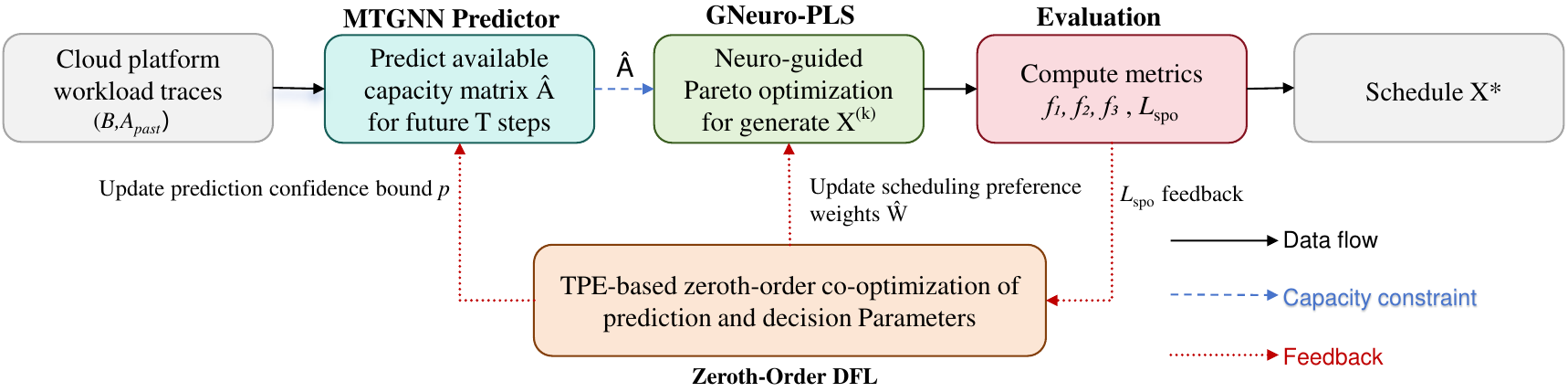} 
    \caption{Overall architecture of the CL-DFL cloud scheduling framework.}
    \label{fig:CL-DFL_framework}
\end{figure*}

Pre-collected task scheduling in cloud platforms inherently constitutes a two-stage PTO 
paradigm under uncertain constraints. To overcome the robustness limitations of traditional decoupled approaches in dynamic environments, the proposed CL-DFL framework establishes a closed-loop interaction between resource forecasting and scheduling optimization. As illustrated in Fig.~\ref{fig:CL-DFL_framework}, CL-DFL integrates perception, decision-making, and evaluation into a unified optimization loop. Specifically, we utilize a zeroth-order DFL mechanism to adaptively update the prediction confidence parameter $p$ and the multi-objective scheduling weights $\hat{\mathbf{W}}$ based on the downstream Pareto performance. This mechanism dynamically calibrates capacity prediction constraints to align predicted resource availability with physical scheduling feasibility. The complete procedure of the E2E zeroth-order CL-DFL framework is presented in Algorithm~\ref{alg:cl_dfl_main}.

\begin{algorithm}[t]
\small 
\SetKwInput{KwIn}{Input}
\SetKwInput{KwOut}{Output}
\DontPrintSemicolon 
\KwIn{Historical capacity $\mathbf{A}_{\text{past}}$, task set $\mathcal{B}$, actual capacity $\mathbf{A}$, target violation threshold $p_{\text{tgt}}$, TPE iterations $K_{\text{BO}}$.}
\KwOut{Optimal control parameters $\{p^*, \hat{\mathbf{W}}^*\}$ and optimal physical scheduling decision matrix $\mathbf{X}^*$.}

\BlankLine
\tcp*[h]{Initialization of CL-DFL and Bayesian surrogate model}\;
Initialize CL-DFL parameters: $p^{(1)}$, $\hat{\mathbf{W}}^{(1)}$, and network parameters $\theta \leftarrow \theta_0$\;
Initialize the TPE surrogate model for black-box decision loss evaluation\;

\BlankLine
\For{$k = 1$ \textbf{to} $K_{\text{BO}}$}{
    The TPE sampler adaptively recommends control parameters $(p^{(k)}, \hat{\mathbf{W}}^{(k)})$\;
    
    \BlankLine
    \tcp*[h]{CL-DFL E2E forward propagation (Alg.~\ref{alg:gtnet_predict} \& Alg.~\ref{alg:gneuro_pls_training})}\;
    $\tilde{\mathbf{A}}, \text{GMM} \leftarrow \text{Predictor}(\mathbf{A}_{\text{past}}, \theta_0)$ \tcc*[r]{Run Alg.~\ref{alg:gtnet_predict}}
    $\bar{a}_{j,t}(p^{(k)}) \leftarrow \tilde{a}_{j,t} + \Phi^{-1}_{\text{GMM}}(p^{(k)}) \quad \forall j, t$ \tcc*[r]{Quantile inverse mapping of GMM to calibrate capacity boundaries}
    $\mathbf{X}^{(k)} \leftarrow \text{Scheduler}(\mathcal{B}, \bar{a}(p^{(k)}), \hat{\mathbf{W}}^{(k)})$ \tcc*[r]{Run Alg.~\ref{alg:gneuro_pls_training}}
    
    \BlankLine
    \tcp*[h]{CL-DFL backward feedback: Physical backtesting and zeroth-order co-tuning}\;
    Evaluate $\mathbf{X}^{(k)}$ under true physical capacity $\mathbf{A}$ to obtain metrics $f_1, f_2, f_3$\;
    $\mathcal{L}_{\text{SPO}}(p^{(k)}, \hat{\mathbf{W}}^{(k)}) \leftarrow \lambda_1 f_1(\mathbf{X}^{(k)}) + \lambda_2 (1 - f_2(\mathbf{X}^{(k)})) + \lambda_3 (1 - f_3(\mathbf{X}^{(k)}))$ \tcc*[r]{Compute SPO decision loss}
    Update the TPE surrogate model using $([p^{(k)}, \hat{\mathbf{W}}^{(k)}], \mathcal{L}_{\text{SPO}})$ to guide subsequent parameter recommendations\;
}

\BlankLine
$\{p^*, \hat{\mathbf{W}}^*\} \leftarrow \arg\min_{p, \hat{\mathbf{W}}} \mathcal{L}_{\text{SPO}}(p, \hat{\mathbf{W}})$\;
\Return $p^*, \hat{\mathbf{W}}^*$ and $\mathbf{X}^*$\;

\caption{E2E Zeroth-Order CL-DFL Cloud Scheduling Framework.}
\label{alg:cl_dfl_main}
\end{algorithm}

\subsection{Spatio-Temporal  Capacity Prediction via MTGNN}
\label{subsec:MTGNN_model}
Accurate capacity forecasting is essential for CL-DFL. We employ MTGNN to capture spatio-temporal dependencies among physical servers and model multi-node workload dynamics.

\subsubsection{Input Representation and Graph Definition}
We represent the historical server-level available capacity observations over the past $T_{\text{in}}$ time steps as a tensor $\mathcal{X} \in \mathbb{R}^{T_{\text{in}} \times S \times d}$, where $S$ denotes the number of servers and $d$ is the feature dimension. The spatial correlations among servers are represented as a graph $\mathcal{G} = (\mathcal{V}, \mathcal{E}, \mathbf{A})$, where $\mathcal{V}$ denotes the set of server nodes and $\mathbf{A} \in \mathbb{R}^{S \times S}$ is the adjacency matrix describing inter-server dependencies.

\subsubsection{Adaptive Graph Learning}
Since workload correlations among physical servers are usually unavailable from prior knowledge, we employ an adaptive graph learning layer to dynamically infer latent spatial dependencies. Given trainable node embedding matrices $\mathbf{E}_1, \mathbf{E}_2 \in \mathbb{R}^{S \times c}$ with dimension $c$, the adaptive adjacency matrix $\mathbf{A}_{\text{adj}}$ is computed as
\begin{equation}
    \mathbf{A}_{\text{adj}} = \text{Softmax}\left(\text{ReLU}\left(\mathbf{E}_1 \mathbf{E}_2^T - \mathbf{E}_2 \mathbf{E}_1^T\right)\right).
\end{equation}
This mechanism enables the model to automatically capture resource synchronization patterns among servers and provides effective spatial representations for capacity prediction.

\subsubsection{Spatio-Temporal Feature Fusion and Uncertainty Calibration}
The MTGNN backbone consists of stacked spatio-temporal layers (ST-Layers), which alternate between temporal convolutional layers (TCLs) and graph convolutional layers (GCLs) to capture workload evolution patterns and inter-server correlations. To characterize the non-stationary and multimodal forecasting errors under heterogeneous workloads, we employ a multi-dimensional Gaussian mixture model (GMM) to model the spatio-temporal residuals $\epsilon_{j,t} = a_{j,t} - \tilde{a}_{j,t}$ collected from the training set, where $a_{j,t}$ and $\tilde{a}_{j,t}$ denote the actual and predicted available capacities of server $j$ at time $t$, respectively. This probabilistic model provides an uncertainty estimation for each server-time pair. By applying the inverse cumulative distribution function (CDF) of the learned residual distribution, $\Phi^{-1}_{\epsilon_{j,t}}(p)$, we derive a dynamically calibrated capacity boundary $\bar{a}_{j,t}(p)$, which is further incorporated into the downstream optimizer as an uncertainty-aware constraint, as detailed in Algorithm~\ref{alg:gtnet_predict}.

\begin{algorithm}[t]
\small
\SetKwInput{KwIn}{Input}
\SetKwInput{KwOut}{Output}
\DontPrintSemicolon
\KwIn{Historical available capacity tensor $\mathcal{X} \in \mathbb{R}^{T_{\text{in}} \times S \times d}$, node embedding matrices $\mathbf{E}_1, \mathbf{E}_2 \in \mathbb{R}^{S \times c}$, MTGNN deep spatio-temporal weight parameters $\mathbf{W}_{k1}, \mathbf{W}_{k2}$.}

\KwOut{Deterministic prediction matrix $\tilde{\mathbf{A}}$ and fitted error distribution $\text{GMM}$.}

\BlankLine
\tcp*[h]{Adaptive graph learning for spatial topology mining}\;
$\mathbf{A}_{\text{adj}} \leftarrow \text{Softmax}\left(\text{ReLU}\left(\mathbf{E}_1 \mathbf{E}_2^T - \mathbf{E}_2 \mathbf{E}_1^T\right)\right)$\;

\BlankLine
\tcp*[h]{Alternating spatio-temporal stacking for deep feature extraction}\;
\For{$l = 1$ \textbf{to} $L$}{
    \tcp*[h]{Capturing temporal dynamic via dilated convolutions}\;
    $\mathcal{X}_{\text{temp}}(t) \leftarrow \sum_{k=0}^{K-1} \mathbf{f}(k) \mathcal{X}(t - d \cdot k)$\;
    
    \BlankLine
    \tcp*[h]{Capturing spatial correlation via graph convolutions}\;
    
    $\mathbf{P}_f \leftarrow \mathbf{A}_{\text{adj}}/\text{rowsum}(\mathbf{A}_{\text{adj}})$\;
    
    $\mathbf{P}_b \leftarrow \mathbf{A}_{\text{adj}}^T/\text{rowsum}(\mathbf{A}_{\text{adj}}^T)$\;

    $\mathcal{Z} \leftarrow \sum_{k=0}^{M} \left( \mathbf{P}_f^k \mathcal{X}_{\text{temp}} \mathbf{W}_{k1} + \mathbf{P}_b^k \mathcal{X}_{\text{temp}} \mathbf{W}_{k2} \right)$\;
    
    Update state: $\mathcal{X} \leftarrow \mathcal{Z}$\;
}

\BlankLine
\tcp*[h]{Prediction output and residual uncertainty distribution fitting}\;
$\tilde{\mathbf{A}} \leftarrow \text{ReLU}(\mathcal{X} \mathbf{W}_{\text{out}} + b_{\text{out}})$\;
Fit historical training error distribution $\epsilon \sim \text{GMM}$\;

\Return $\tilde{\mathbf{A}}$ and $\text{GMM}$\;

\caption{MTGNN Multi-Node Capacity Prediction.}
\label{alg:gtnet_predict}
\end{algorithm}

\subsection{Multi-Objective Scheduling Algorithm Based on Neural Pareto Local Search}
\label{subsec:neuro_pls_scheduler}
Pre-collected task scheduling in cloud platforms
involves large discrete search spaces, tightly coupled capacity and temporal constraints, and inherent conflicts among violation rate, user satisfaction, and resource utilization.  To address these challenges, we adopt neural Pareto local search (Neuro-PLS)~\cite{zhang2025neuropls} as the underlying multi-objective optimization framework. Neuro-PLS decomposes the Pareto optimization process into a sequence of scalarized sub-problems and leverages deep neural networks to guide neighborhood exploration toward promising solutions. Within the CL-DFL framework, Neuro-PLS provides an effective optimization backbone for searching Pareto-optimal scheduling strategies under complex cloud resource constraints.

\subsubsection{MOCOP  Modeling}
For any scheme $\mathbf{X} \in \Omega$, the multi-objective evaluation function $\mathbf{F}(\mathbf{X})$ is defined as
\begin{equation}
    \min \mathbf{F}(\mathbf{X})=[f_1(\mathbf{X}), 1-f_2(\mathbf{X}), 1-f_3(\mathbf{X})]^T.
\end{equation}
This formulation characterizes the inherent trade-offs among resource safety, QoS requirements, and system efficiency.

\subsubsection{Scalarization via Decomposition}
To efficiently explore the Pareto front, Neuro-PLS adopts the decomposition strategy of MOEA/D \cite{zhang2007moea}. Specifically, a set of uniformly distributed weight vectors $\hat{\mathbf{W}} = \{\mathbf{w}^1, \mathbf{w}^2, \dots, \mathbf{w}^H\}$ is generated, where each vector $\mathbf{w}^i = (w_1^i, w_2^i, w_3^i)$ represents a different preference safety, QoS, and efficiency. For each weight vector, the Tchebycheff approach is utilized to evaluate the optimization potential of solution $\mathbf{X}$ with respect to the reference point $\mathbf{z}^*$:
\begin{equation}
    g^{te}(\mathbf{X} \mid \mathbf{w}^i, \mathbf{z}^*) = \max_{j \in \{1,2,3\}} \{ w_j^i \cdot |F_j(\mathbf{X}) - z_j^*| \}.
\end{equation}
By minimizing $g^{te}$ under different weight preferences, Neuro-PLS decomposes the original MOCOP into multiple scalar sub-problems and progressively approximates the Pareto-optimal boundary.

\subsubsection{Neuro-Guided Seed Solution Selection and Adaptive Search}
To improve the efficiency of local search, Neuro-PLS employs a scoring network $\psi$ to evaluate candidate seed solutions. Specifically, given the solution feature vector $v_i$ of scheme $x_i$, the network estimates its search potential through nonlinear scoring and selects a promising seed solution as the search center, thereby reducing random exploration. Furthermore, a neural adaptive first-$K$ search intensity control strategy is designed to balance global exploration and local exploitation.
This mechanism adaptively adjusts the neighborhood search radius according to the available computational budget. This mechanism enables wider exploration in early iterations and progressively focuses on high-quality regions of the Pareto front, improving both search efficiency and convergence performance.

\subsection{Zeroth-Order Cooperative Policy Evolution via GRPO}
\label{subsec:grpo_training_details}
Since Neuro-PLS operates on a discrete black-box MOCOP, explicit gradients with respect to policy parameters are unavailable. To overcome this, we develop GNeuro-PLS, a zeroth-order cooperative policy optimization framework that integrates GRPO with parameter perturbation-based evolution, enabling policy adaptation without gradient computation.

\subsubsection{Markov Decision Process Formulation}
We reformulate the adaptive local search process as a multi-agent Markov decision process (MDP). 
The state is defined as $s_t=(\rho_t,S'_t)$, where $\rho_t=FE_t/Budget$ denotes the consumed evaluation budget ratio and $S'_t$ represents the current non-dominated solution archive. The action is jointly determined by two cooperative agents: a GNN-based agent that selects a promising seed solution $\mathbf{X}^*\in S'_t$, and a multi-layer perceptron (MLP)-based agent that adaptively adjusts the neighborhood truncation radius ($K$). The reward is obtained by directly evaluating the generated scheduling solution $\mathbf{X}$ under the actual physical capacity: 
\begin{equation}
    r_t = -\mathcal{L}_{\text{SPO}} = -\left(\lambda_1 f_1 + \lambda_2 (1-f_2) + \lambda_3 (1-f_3) \right),
\end{equation}
where $r_t$ represents the reward at step $t$, $\mathcal{L}_{\text{SPO}}$ is the comprehensive smart predict-and-optimize (SPO) decision loss, and $\lambda_1$, $\lambda_2$, and $\lambda_3$ denote balance system safety, user experience, and resource efficiency, respectively.
\subsubsection{Dual-Mirror Sampling and Group Relative Advantage}
For each input state, we generate $G = 2H$ trajectories by symmetric policy perturbations and perform intra-group relative comparisons to estimate zeroth-order optimization directions. Specifically, for each Gaussian noise vector $\mathbf{\epsilon}_g \sim \mathcal{N}(0, \mathbf{I})$, the baseline policy parameters $\theta$ are symmetrically perturbed as
\begin{equation}
    \theta_+^{(g)} = \theta + \sigma \epsilon_g, \quad \theta_-^{(g)} = \theta - \sigma \epsilon_g, \quad g \in \{1, \dots, H\},
\end{equation}
where $\sigma$ denotes the perturbation step size. Let
$\mathbf{r} = [r_+^{(1)}, r_-^{(1)}, \dots, r_+^{(H)}, r_-^{(H)}]$ denote the collected rewards from all perturbed policies. Following the GRPO paradigm, we compute the normalized advantage $A_i$ for each reward $r_i$ as 
\begin{equation}
    A_i = \frac{r_i - \text{mean}(\mathbf{r})}{\text{std}(\mathbf{r}) + \epsilon_{0}},
\end{equation}
where $\mathbf{I}$ is the identity matrix, $\text{mean}(\mathbf{r})$ and $\text{std}(\mathbf{r})$ denote the group average and group standard deviation of the reward set $\mathbf{r}$, respectively, and $\epsilon_0$ is a small constant for numerical stability. This relative normalization eliminates the need for an explicit critic network and improves training robustness under heterogeneous workload uncertainties.
\subsubsection{Zeroth-Order Gradient Adaptive Update}
By exploiting the advantage differences between dual-mirror perturbations, we estimate the zeroth-order gradient direction and adaptively update the GNeuro-PLS policy parameters as
\begin{equation}
    \theta \leftarrow \theta + \frac{\alpha}{H \sigma} \sum_{g=1}^H \left( A_{2g-1} - A_{2g} \right) \epsilon_g,
\end{equation}
where $\alpha$ is the adaptive evolutionary learning rate and $H$ denotes the number of dual-mirror perturbation pairs ($H = G/2$).
This update mechanism utilizes intra-group relative advantages derived from scheduling feedback, enabling efficient policy evolution without requiring explicit gradient information.

\subsubsection{Algorithmic Flow}
By integrating multi-objective scheduling with zeroth-order GRPO, we formalize the adaptive resource orchestration procedure, as presented in Algorithms~\ref{alg:gneuro_pls_training} and Algorithm~\ref{alg:gneuro_pls_inference}. Following an offline training and online inference paradigm, CL-DFL effectively balances conflicting objectives, i.e., violation rate, user satisfaction, and resource utilization, within high-dimensional and non-convex solution spaces. Meanwhile, it achieves near-real-time scheduling with second-level inference latency.

\begin{algorithm}[t]
\small
\SetKwInput{KwIn}{Input}
\SetKwInput{KwOut}{Output}
\DontPrintSemicolon 
\KwIn{Task set $\mathcal{B}$, predicted capacity $\hat{\mathbf{A}}$, physical capacity $\mathbf{A}$, initial parameters $\theta_{\text{GNN}}, \theta_{\text{MLP}}$, group size $G$, training epochs $E_{\text{MLP}}$, $E_{\text{GNN}}$.}
\KwOut{Optimal policy weights $\theta_{\text{MLP}}^*$ and $\theta_{\text{GNN}}^*$.}

\BlankLine
\tcp*[h]{MLP Policy Evolutionary Training}\;
\For{$e = 1$ \textbf{to} $E_{\text{MLP}}$}{
    $\theta_{\text{MLP}}^{(g)} = \theta_{\text{MLP}} + \sigma \epsilon_g \quad (\epsilon_g \sim \mathcal{N}(0, \mathbf{I}), \; g \in \{1,\dots,G\})$\;
    \For{$g = 1$ \textbf{to} $G$}{
        $\mathbf{X}_g \leftarrow \text{GNeuro-PLS-Random}(\mathcal{B}, \hat{\mathbf{A}}, \theta_{\text{MLP}}^{(g)})$\;
        $r_g \leftarrow -\mathcal{L}_{\text{SPO}}(\mathbf{X}_g, \mathbf{A})$\;
    }
    $A_g = \frac{r_g - \text{mean}(\mathbf{r})}{\text{std}(\mathbf{r}) + \epsilon_{0}}$\;
    $\theta_{\text{MLP}} \leftarrow \theta_{\text{MLP}} + \frac{\alpha_{\text{MLP}}}{G \sigma} \sum_{g=1}^G A_g \epsilon_g$\;
}
$\theta_{\text{MLP}}^* \leftarrow \theta_{\text{MLP}}$\;

\BlankLine
\tcp*[h]{GNN Policy Evolutionary Training}\;
\For{$e = 1$ \textbf{to} $E_{\text{GNN}}$}{
    $\theta_{\text{GNN}}^{(g)} = \theta_{\text{GNN}} + \sigma \epsilon_g \quad (\epsilon_g \sim \mathcal{N}(0, \mathbf{I}), \; g \in \{1,\dots,G\})$\;
    \For{$g = 1$ \textbf{to} $G$}{
        $\mathbf{X}_g \leftarrow \text{GNeuro-PLS}(\mathcal{B}, \hat{\mathbf{A}}, \theta_{\text{GNN}}^{(g)}, \theta_{\text{MLP}}^*)$\;
        $r_g \leftarrow -\mathcal{L}_{\text{SPO}}(\mathbf{X}_g, \mathbf{A})$\;
    }
    $A_g = \frac{r_g - \text{mean}(\mathbf{r})}{\text{std}(\mathbf{r}) + \epsilon_{0}}$\;
    $\theta_{\text{GNN}} \leftarrow \theta_{\text{GNN}} + \frac{\alpha_{\text{GNN}}}{G \sigma} \sum_{g=1}^G A_g \epsilon_g$\;
}
$\theta_{\text{GNN}}^* \leftarrow \theta_{\text{GNN}}$\;

\BlankLine
\Return $\theta_{\text{MLP}}^*$ and $\theta_{\text{GNN}}^*$\;

\caption{GNeuro-PLS Offline Policy Training.}
\label{alg:gneuro_pls_training}
\end{algorithm}

\begin{algorithm}[t]
\small
\SetKwInput{KwIn}{Input}
\SetKwInput{KwOut}{Output}
\DontPrintSemicolon
\KwIn{Task set $\mathcal{B}$, predicted capacity $\hat{\mathbf{A}}$, trained optimal parameters $\theta_{\text{GNN}}^*, \theta_{\text{MLP}}^*$, evaluation budget $\text{\emph{Budget}}$.}
\KwOut{Optimal scheduling decision matrix set $\mathbf{X}^*$ (Pareto front).}

\BlankLine
Initialize Pareto archive $\mathcal{A} \leftarrow \text{InitRandomSolutions()}$, reset counter $\text{FE} \leftarrow 0$\;

\While{$\text{FE} < \text{\emph{Budget}}$ \textbf{and} there exist unexplored schemes in $\mathcal{A}$}{
    \tcp*[h]{GNN Selects Seed}\;
    $x^* \leftarrow \arg\max_{x \in \mathcal{A}} \text{GNN}(x, \theta_{\text{GNN}}^*)$\;
    
    \BlankLine
    \tcp*[h]{MLP Determines Step Size}\;
    $K \leftarrow \text{MLP}(\text{progress}, \theta_{\text{MLP}}^*)$\;
    
    \BlankLine
    \tcp*[h]{Neighborhood Exploration}\;
    Reschedule $K$ tasks in $x^*$ to generate candidate $x'$\;
    
    \BlankLine
    \tcp*[h]{Pareto Filtering}\;
    Evaluate physical metrics of $x'$ and increment $\text{FE} \leftarrow \text{FE} + 1$\;
    
    \textbf{if} $x'$ is not dominated by any existing solution in $\mathcal{A}$\; \textbf{then}\;
    \Indp
        Merge $x'$ into $\mathcal{A}$ and prune dominated solutions\;
    \Indp
}

\BlankLine
$\mathbf{X}^* \leftarrow \mathcal{A}$\;
\Return Final non-dominated scheduling scheme set $\mathbf{X}^*$\;

\caption{GNeuro-PLS Online Scheduling Inference.}
\label{alg:gneuro_pls_inference}
\end{algorithm}

\subsection{TPE-Based Zeroth-Order DFL Optimization}
\label{subsec:tpe_dfl_loop}
Traditional cloud resource management usually separates workload forecasting from downstream scheduling. This disconnection often leads to prediction models to optimize surrogate statistical objectives (e.g., MSE) rather than decision-oriented performance. To address this, we develop a zeroth-order DFL mechanism that establishes a closed-loop feedback pathway between scheduling outcomes and prediction-oriented constraints.

In uncertain cloud environments, true physical capacity constraints $\mathbf{A}$ are often difficult to obtain accurately in advance. Therefore, we transform the predicted capacity distribution $\hat{\mathbf{A}}$ into uncertainty-aware constraint boundaries $\bar{a}_{j,t}(p)$ 
through quantile calibration, where the confidence level $p \in (0, 1)$ controls the conservativeness of resource constraints. Given the calibrated boundary, the downstream scheduling problem is formulated as
\begin{equation}
    \mathbf{X}^*(p, \hat{\mathbf{W}}) = \arg\min_{\mathbf{X} \in \Omega} \hat{\mathbf{W}}^T \mathbf{F}(\mathbf{X}) \quad \text{s.t.} \quad u_{j,t} \le \bar{a}_{j,t}(p).
\end{equation}
After execution, the scheduling outcome is evaluated using the true physical capacity, producing the SPO decision loss $\mathcal{L}_{\text{SPO}}$. Since GNeuro-PLS is a non-differentiable black-box optimizer, the analytical gradient $\nabla_p \mathcal{L}_{\text{SPO}}$ is unavailable, making conventional gradient-based DFL infeasible.

To tackle this, we employ a Bayesian optimizer based on the TPE as a zeroth-order DFL solver. Instead of updating the forecaster parameters $\theta$, TPE directly optimizes decision-facing parameters $\Theta =\{p, \hat{\mathbf{W}}\}$ by leveraging historical scheduling feedback: 
\begin{equation}
    \Theta^* = \{p^*, \hat{\mathbf{W}}^*\} = \arg\min_{p, \hat{\mathbf{W}}} \mathcal{L}_{\text{SPO}}(p, \hat{\mathbf{W}}).
\end{equation}

Unlike gradient-based DFL methods that require computationally expensive network fine-tuning, CL-DFL freezes the pretrained forecaster at $\theta_0$ and adaptively calibrates uncertainty boundaries and objective preferences. This design avoids the non-differentiability of combinatorial schedulers while preserving second-level online inference efficiency. Through the closed-loop feedback mechanism, CL-DFL aligns prediction uncertainty with actual scheduling requirements, reducing decision errors caused by forecasting bias and improving E2E cloud orchestration performance.
\section{Performance Evaluation}
This section evaluates the proposed CL-DFL framework using large-scale real-world cloud workload traces. We analyze its convergence, Pareto optimization, and scheduling performance under heterogeneous workloads.
Experimental comparisons validate the advantages of the proposed framework in mitigating violation rates, improving user satisfaction, and enhancing resource utilization.

\subsection{Experimental Setup and Metrics}

\subsubsection{Datasets}
We select four public  workload datasets from Microsoft Azure and Alibaba, and reconstruct them into fine-grained spatio-temporal workloads for server-level capacity forecasting and scheduling. Experiments are conducted under three workload scales (140, 220, and 300 tasks) to evaluate the performance of CL-DFL under different workload intensities.

\textbf{Microsoft Azure Large Model Inference Trace Datasets (2023~/~2024~/~2025)}. 
Our experiments cover the 2023 dataset ~\cite{patel2024splitwise}, 2024~\cite{stojkovic2025dynamollm}, and 2025 multi-modal LLM inference traces ~\cite{qiu2026modserve}. We extract system capacity at an hourly time step and construct fine-grained spatio-temporal samples of available capacity matrices through a sliding-window mechanism. Meanwhile, pre-collected task requests are sampled from traces, where CPU demands, execution durations, and temporal flexibility windows are estimated according to workload characteristics, including ContextTokens and GeneratedTokens.

\textbf{Alibaba Industrial-Grade Co-located Cluster Trace Dataset (2018)}. The Alibaba Cluster-trace-2018 dataset \cite{zhu2023} records co-located deployments of online containerized microservices and offline batch jobs across thousands of physical machines over an eight-day period.
The coexistence of heterogeneous workloads results in dynamic resource demands and fluctuating available capacity, providing a representative environment for evaluating pre-collected job scheduling. We incorporate this dataset to further evaluate the generalization capability of CL-DFL in realistic multi-workload cloud orchestration scenarios.

\subsubsection{Baseline Methods}
We select five representative algorithms for comparison: \emph{CUC}~\cite{dong2021predictive},  a PTO baseline that employs resource forecasting with predefined safety margins; \emph{NSGA-II}~\cite{deb2002fast}, a classical multi-objective evolutionary algorithm widely used for Pareto optimization; \emph{PLS}~\cite{lin2022pareto}, a heuristic Pareto local search method for discrete scheduling; \emph{SEMO}~\cite{laumanns2004running}, a lightweight evolutionary method based on single-solution mutation; \emph{Neuro-PLS}~\cite{zhang2025neuropls}, a neural-guided PLS variant without GRPO, used to evaluate the contribution of the proposed cooperative scheduling mechanism.






To ensure fairness, all baseline hyperparameters (e.g., the static safety margin in CUC) are tuned using the same TPE-based validation procedure, and each baseline is evaluated with its best validation configuration under identical testing settings.

\subsubsection{Experimental Settings}
 
Experiments are conducted on a workstation with an AMD Threadripper 3970X CPU and an NVIDIA RTX 3090 Ti GPU. The framework is implemented in Python 3.8 with PyTorch, while the scheduling module is accelerated by C++ to achieve second-level inference latency. All baseline methods are implemented and evaluated under the same environment for fair comparison.

\textbf{Evaluation Protocol}. To avoid data leakage, datasets are chronologically divided into training into training (60\%), validation (20\%), and testing (20\%) sets. The outer-loop TPE optimization is performed only on the validation set to determine the DFL parameters, including $p$, $\hat{\mathbf{W}}$, and $t_{\text{delay}}$. The selected optimal parameters 
 $\{p^*, \hat{\mathbf{W}}^*\}$ are then fixed and used for final evaluation on the unseen test set. All reported results, including baseline comparisons and ablation studies, are averaged over five independent runs to reduce stochastic variations.
 

\textbf{Hyperparameter Selection}. Table~\ref{tab:hyperparameters_list} summarizes the selected hyperparameters of CL-DFL. The outer-loop TPE procedure optimizes the DFL control parameters on the validation set, which are then fixed for final testing. Under Azure workloads with 140, 220, and 300 tasks, $\{p^*, t_{\text{delay}}, \hat{\mathbf{W}}^*\}$ remain stable at 0.8286, 3 hours, and (0.3276, 0.3940, 0.2785), respectively. The GNeuro-PLS scheduler is evaluated with a search budget of 200,000 steps under identical settings.

\begin{table}[ht]
\centering
\caption{Hyperparameter Configurations Across CL-DFL Sub-modules.}
\label{tab:hyperparameters_list}
\renewcommand{\arraystretch}{1.05}
\setlength{\tabcolsep}{3.5pt} 
\begin{tabular}{p{4.2cm} c c}
\toprule
\textbf{Hyperparameter Description} & \textbf{Symbol} & \textbf{Value} \\ \midrule
Historical history input length (h) & $T_{\text{in}}$ & 120 \\
Number of physical servers & $S$ & 10 \\
Future forecasting horizon (h) & $T$ & 24 \\
Node feature embedding dimension & $c$ & 10 \\
Max spatial diffusion steps & $M$ & 2 \\
Spatiotemporal layer count & $l$ & 3 \\ \midrule
Bayesian optimization epochs & $K_{\text{BO}}$ & 50 \\
Confidence quantile bounds & $p$ & (0.60, 1.00) \\
Delay tolerance bounds (h) & $t_{\text{delay}}$ & [1.0, 3.5] \\
SLA/Experience/Energy weights & $\lambda_{1:3}$ & 0.5, 0.3, 0.2 \\ \midrule
Tchebycheff sub-direction weights & $H$ & 15 \\
Local search evaluation budget & $Budget$ & 200,000 \\ \midrule
Co-evolutionary strategy epochs & $E$ & 250 \\
Dual-mirror perturbation size & $G$ & 16 \\
Perturbation scale parameter & $\sigma$ & 0.05 \\
Policy evolutionary learning rate & $\alpha$ & 0.02 \\
\bottomrule
\end{tabular}
\end{table}

\subsection{Performance Comparison}
Utilizing heterogeneous workload traces from Azure (2023–2025) and Alibaba cluster trace (2018), we evaluate CL-DFL in terms of convergence behavior and multi-objective scheduling performance, including  violation rate, user satisfaction, and resource utilization.
\subsubsection{Convergence Analysis}
\begin{figure}[h]
    \centering   \includegraphics[width=0.75\linewidth]{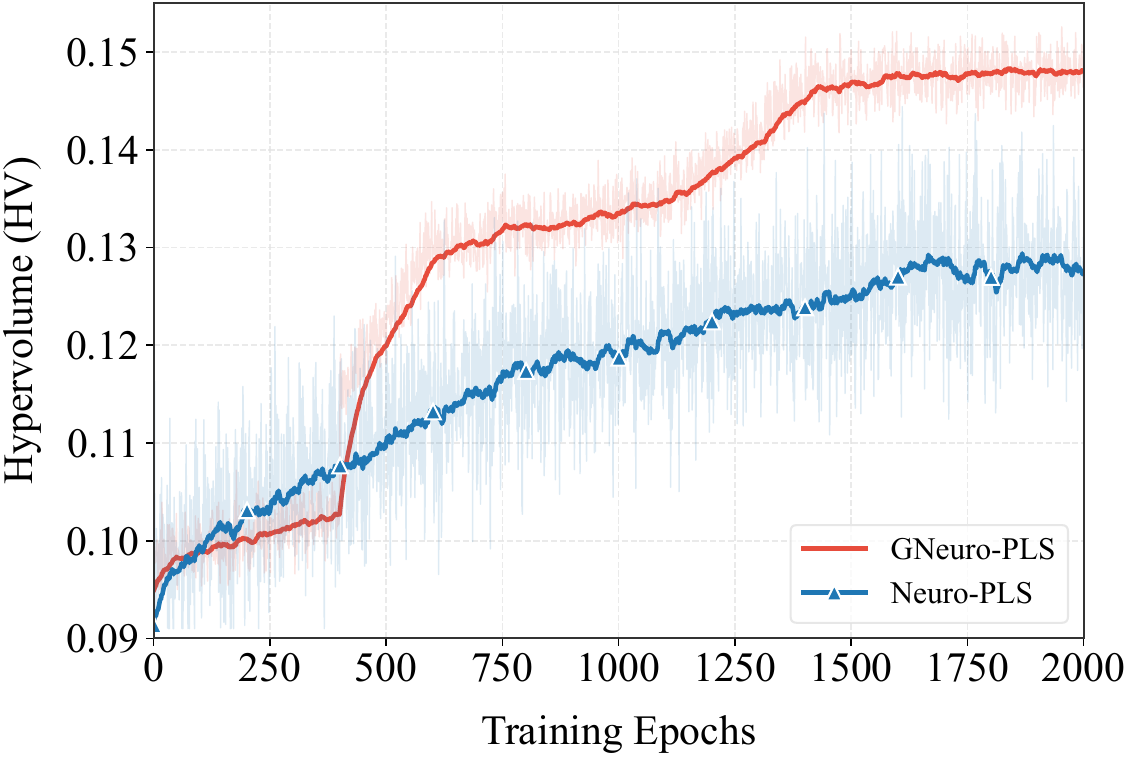}
    \caption{HV convergence curves of GNeuro-PLS vs. Neuro-PLS.}
    \label{fig:hv_convergence}
\end{figure}
Fig.~\ref{fig:hv_convergence} 
presents the Hypervolume (HV) \cite{Shang2021} convergence curves of GNeuro-PLS and Neuro-PLS under high-pressure workloads with 600 tasks, where HV measures the dominated objective-space volume (larger is better) and reflects the convergence and diversity of Pareto solutions. During the early training stage (0–400 epochs), GNeuro-PLS achieves a slightly lower HV value due to the constrained exploration budget introduced by first-$K$ search intensity control
strategy. However, this stage enables the model to gradually learn effective heuristic guidance. As the exploration budget increases, GNeuro-PLS exhibits rapid HV improvements around the 400th and 1100th epochs, indicating enhanced search capability in complex discrete solution spaces. In contrast, Neuro-PLS without the GRPO-based cooperative mechanism converges prematurely and suffers from stagnation after approximately 1500 epochs. By leveraging intra-group relative advantage normalization, GNeuro-PLS achieves a higher final HV value (approximately 0.148) with smaller fluctuations during training, demonstrating improved convergence stability and robustness under heterogeneous workload environments.


\subsubsection{Scheduling Efficiency and QoS Compliance}
We evaluate the scheduling performance of CL-DFL from three perspectives: violation rate, user satisfaction, and decision-making efficiency.

\textbf{Violation Rate under Uncertain Capacity}. Experimental results demonstrate the robustness of CL-DFL under uncertain capacity estimation. As shown in Table~\ref{tab:comparison_scale_140}, under light workloads (140 tasks), CL-DFL maintains the SLA violation rate below 0.5$\%$, significantly outperforming the CUC baseline (3.50$\%$–5.80$\%$). 
When the workload scale increases to 220 tasks, as illustrated in Table~\ref{tab:comparison_scale_220}, CL-DFL consistently keeps the violation rate within the predefined  5$\%$ threshold on Azure traces, whereas Neuro-PLS without the dual-agent policy selection mechanism suffers from violation rates exceeding 10$\%$. This result indicates that the TPE-based parameter adaptation effectively improves scheduling robustness by adjusting the uncertainty-related confidence parameter $p$. Under the heterogeneous Alibaba 2018 co-located workload with 300 tasks, as shown in Table~\ref{tab:comparison_scale_300}, all methods experience increased violations due to intensified resource contention. Nevertheless, CL-DFL achieves a lower violation rate of 25.83$\%$, compared with 30.36$\%$ for PLS, demonstrating its superior robustness under complex workload conditions.

\textbf{User Experience and Resource Utilization Trade-off}. 
We further evaluate the trade-off between user satisfaction and resource utilization in the multi-objective scheduling problem. Across all workload scales of the Azure traces, CL-DFL maintains a user satisfaction rate ($f_2$) above 99.8$\%$, demonstrating its capability to satisfy diverse temporal requirements. Under the heterogeneous Alibaba 2018 co-located workload, the CUC baseline suffers a significant degradation in user satisfaction, with the rate decreasing 46.04$\%$, due to its limited capability in handling conflicting objectives. In contrast, CL-DFL preserves nearly full user satisfaction while achieving effective resource allocation. Moreover, under heavy workloads with 300 tasks, CL-DFL achieves a resource utilization rate ($f_3$) exceeding 80$\%$, substantially outperforming NSGA-II (35$\%$-52$\%$). These results demonstrate that CL-DFL achieves a favorable balance between QoS satisfaction and resource utilization under complex heterogeneous cloud workloads.

\textbf{Real-Time Decision-Making Efficiency}.
We further evaluate the decision-making latency of different scheduling approaches under increasing workload scales. As the number of tasks increases from 140 to 300, traditional evolutionary algorithms (NSGA-II and SEMO) incur rapidly increasing search overhead, resulting in minute-level scheduling latency (90–140 s). Meanwhile, CUC reaches up to 305 s under 220 tasks, limiting its applicability to dynamic cloud orchestration scenarios. In contrast, CL-DFL achieves substantially lower decision latency, requiring less than 1.1 s on heavy Azure workloads and approximately 20 s on the Alibaba 2018 trace. Although the maximum search budget of GNeuro-PLS is configured as 200,000 iterations, the GNN-guided initialization and adaptive search control enable early termination after only several thousand effective evaluations (typically 2,000–5,000 steps). This significantly reduces unnecessary exploration while maintaining solution quality. The results demonstrate that CL-DFL effectively balances optimization performance and computational efficiency, enabling low-latency scheduling decisions for heterogeneous cloud workloads.


\begin{table*}[h]
\centering
\scriptsize
\setlength{\tabcolsep}{3pt}
\renewcommand{\arraystretch}{1.2}
\caption{Performance comparison of various scheduling algorithms across four real-world industrial datasets (140 tasks).}
\label{tab:comparison_scale_140}
\begin{tabular}{l cccc cccc cccc cccc}
\toprule
\multirow{2}{*}{Method} & \multicolumn{4}{c}{\textbf{Azure 2023}} & \multicolumn{4}{c}{\textbf{Azure 2024}} & \multicolumn{4}{c}{\textbf{Azure 2025}} & \multicolumn{4}{c}{\textbf{Alibaba 2018}} \\
\cmidrule(lr){2-5} \cmidrule(lr){6-9} \cmidrule(lr){10-13} \cmidrule(lr){14-17}
 & $f_1\downarrow$ & $f_2\uparrow$ & $f_3\uparrow$ & Time(s)$\downarrow$ & $f_1\downarrow$ & $f_2\uparrow$ & $f_3\uparrow$ & Time(s)$\downarrow$ & $f_1\downarrow$ & $f_2\uparrow$ & $f_3\uparrow$ & Time(s)$\downarrow$ & $f_1\downarrow$ & $f_2\uparrow$ & $f_3\uparrow$ & Time(s)$\downarrow$ \\
\midrule
CUC (Baseline) & \underline{3.50} & 76.50 & 37.76 & \textbf{2.54} & 2.92 & 74.57 & 38.95 & \textbf{2.39} & 5.80 & 70.79 & \textbf{62.35} & 41.24 & \textbf{1.26} & 55.16 & 32.78 & 2.74 \\
NSGA-II      & 10.56 & 98.23 & 33.81 & 60.44 & 12.68 & \underline{98.30} & 37.14 & 90.24 & 22.42 & 96.45 & 52.33 & 58.99 & 7.97 & 90.27 & 26.86 & 71.68 \\
PLS          & 5.51 & 66.51 & 32.18 & 69.28 & 9.83 & 71.22 & 38.71 & 71.70 & 47.50 & 56.30 & 58.72 & 71.22 & \underline{1.99} & 50.41 & \underline{33.48} & 56.60 \\
SEMO         & \textbf{0.00} & 98.91 & \underline{41.16} & 80.85 & \textbf{0.00} & \textbf{100.00} & \textbf{46.29} & 139.43 & 3.75 & 81.43 & \underline{58.88} & 60.69 & 14.98 & 85.75 & \textbf{37.35} & 60.69 \\
Neuro-PLS    & \textbf{0.00} & \textbf{100.00} & 39.97 & 7.75 & \textbf{0.00} & \textbf{100.00} & 41.10 & \underline{10.06} & \textbf{0.00} & \textbf{99.93} & 41.79 & \underline{10.13} & 14.58 & \underline{99.46} & 29.05 & \underline{2.13} \\
CL-DFL (Ours) & \textbf{0.00} & \underline{99.86} & \textbf{41.82} & \underline{5.37} & \underline{0.17} & \textbf{100.00} & \underline{44.59} & 10.82 & \underline{0.50} & \underline{99.86} & 41.66 & \textbf{8.83} & 12.08 & \textbf{100.00} & 30.23 & \textbf{0.96} \\
\bottomrule
\end{tabular}
\par\smallskip
\noindent \textbf{Note}: Boldface indicates the optimal value; underline denotes the suboptimal value. The values of $f_1$, $f_2$, and $f_3$ are presented in percentage (\%).
\end{table*}

\begin{table*}[h]
\centering
\scriptsize
\setlength{\tabcolsep}{3pt}
\renewcommand{\arraystretch}{1.2}
\caption{Performance comparison of various scheduling algorithms across four real-world industrial datasets (220 tasks).}
\label{tab:comparison_scale_220}
\begin{tabular}{l cccc cccc cccc cccc}
\toprule
\multirow{2}{*}{Method} & \multicolumn{4}{c}{\textbf{Azure 2023}} & \multicolumn{4}{c}{\textbf{Azure 2024}} & \multicolumn{4}{c}{\textbf{Azure 2025}} & \multicolumn{4}{c}{\textbf{Alibaba 2018}} \\
\cmidrule(lr){2-5} \cmidrule(lr){6-9} \cmidrule(lr){10-13} \cmidrule(lr){14-17}
 & $f_1\downarrow$ & $f_2\uparrow$ & $f_3\uparrow$ & Time(s)$\downarrow$ & $f_1\downarrow$ & $f_2\uparrow$ & $f_3\uparrow$ & Time(s)$\downarrow$ & $f_1\downarrow$ & $f_2\uparrow$ & $f_3\uparrow$ & Time(s)$\downarrow$ & $f_1\downarrow$ & $f_2\uparrow$ & $f_3\uparrow$ & Time(s)$\downarrow$ \\
\midrule
CUC (Baseline) & 5.71 & 76.22 & \underline{57.40} & 4.05 & \textbf{2.42} & 77.84 & 61.73 & 305.39 & \underline{7.00} & 58.21 & 69.41 & 42.77 & \textbf{4.17} & 53.57 & \textbf{54.76} & \textbf{12.01} \\
NSGA-II & 15.88 & 96.54 & 41.21 & 84.47 & 18.79 & 97.39 & 44.45 & 120.30 & 26.19 & 98.40 & 66.73 & 86.97 & 13.14 & 87.49 & 35.21 & 117.13 \\
PLS & 14.92 & 67.52 & 46.89 & 70.84 & 21.60 & 72.09 & 52.08 & 91.53 & 58.70 & 59.46 & \underline{72.01} & 79.40 & \underline{9.76} & 51.17 & 39.14 & 83.12 \\
SEMO & \textbf{3.05} & 95.84 & 50.39 & 79.06 & 9.45 & 91.23 & 64.31 & 104.10 & 41.45 & 76.52 & \textbf{79.24} & 85.35 & 18.35 & 59.64 & \underline{44.35} & 112.51 \\
Neuro-PLS & 14.75 & \textbf{100.00} & \textbf{57.61} & \underline{1.55} & 3.67 & \textbf{100.00} & \underline{68.45} & \underline{2.04} & 10.08 & \underline{99.74} & 60.31 & \underline{3.02} & 22.08 & \underline{99.78} & 33.04 & \underline{14.48} \\
CL-DFL (Ours) & \underline{3.92} & \textbf{100.00} & 56.54 & \textbf{1.42} & \underline{3.33} & \underline{99.99} & \textbf{68.55} & \textbf{1.45} & \textbf{2.75} & \textbf{100.00} & 63.77 & \textbf{3.00} & 20.92 & \textbf{99.95} & 36.59 & 16.12 \\
\bottomrule
\end{tabular}
\par\smallskip
\end{table*}


\begin{table*}[h]
\centering
\scriptsize
\setlength{\tabcolsep}{3pt}
\renewcommand{\arraystretch}{1.2}
\caption{Performance comparison of various scheduling algorithms across four real-world industrial datasets (300 tasks).}
\label{tab:comparison_scale_300}
\begin{tabular}{l cccc cccc cccc cccc}
\toprule
\multirow{2}{*}{Method} & \multicolumn{4}{c}{\textbf{Azure 2023}} & \multicolumn{4}{c}{\textbf{Azure 2024}} & \multicolumn{4}{c}{\textbf{Azure 2025}} & \multicolumn{4}{c}{\textbf{Alibaba 2018}} \\
\cmidrule(lr){2-5} \cmidrule(lr){6-9} \cmidrule(lr){10-13} \cmidrule(lr){14-17}
 & $f_1\downarrow$ & $f_2\uparrow$ & $f_3\uparrow$ & Time(s)$\downarrow$ & $f_1\downarrow$ & $f_2\uparrow$ & $f_3\uparrow$ & Time(s)$\downarrow$ & $f_1\downarrow$ & $f_2\uparrow$ & $f_3\uparrow$ & Time(s)$\downarrow$ & $f_1\downarrow$ & $f_2\uparrow$ & $f_3\uparrow$ & Time(s)$\downarrow$ \\
\midrule
CUC (Baseline) & 33.25 & 77.17 & \underline{69.20} & 41.63 & \textbf{3.00} & 70.83 & 69.29 & 209.74 & \textbf{0.83} & 37.73 & 64.52 & 42.50 & \textbf{2.03} & 46.04 & \textbf{68.75} & 198.21 \\
NSGA-II & \underline{22.06} & 95.64 & 48.59 & 112.67 & 23.10 & 98.10 & 51.17 & 120.08 & 39.56 & 97.84 & 72.50 & 112.73 & \underline{21.23} & 85.31 & 41.22 & 146.82 \\
PLS & 28.03 & 65.79 & 57.92 & 68.84 & 34.14 & 66.11 & 59.24 & 77.38 & 57.30 & 62.58 & 78.58 & 81.71 & 30.36 & 44.55 & 54.51 & 116.23 \\
SEMO & \textbf{15.97} & 83.48 & 68.33 & 79.08 & 22.44 & 86.08 & 74.01 & 85.07 & 43.75 & 75.07 & 76.73 & 104.89 & 22.31 & 57.87 & \underline{54.90} & 140.71 \\
Neuro-PLS & 33.00 & \textbf{99.99} & 67.75 & \textbf{0.76} & 22.67 & \underline{99.87} & \underline{81.98} & \textbf{0.97} & 20.42 & \underline{99.89} & \underline{80.08} & \underline{0.75} & 25.42 & \underline{97.45} & 37.76 & \underline{29.77} \\
CL-DFL (Ours) & 28.58 & \underline{99.98} & \textbf{73.13} & \underline{0.80} & \underline{21.75} & \textbf{99.91} & \textbf{84.18} & \underline{1.08} & \underline{18.50} & \textbf{99.96} & \textbf{81.98} & \textbf{0.63} & 25.83 & \textbf{99.93} & 42.58 & \textbf{19.89} \\
\bottomrule
\end{tabular}
\par\smallskip
\end{table*}


\subsection{Ablation Study and Robustness Analysis}
To investigate the contribution of individual components in CL-DFL, we conduct ablation experiments under three workload intensities. Specifically, we compare CL-DFL with DPTO-Naive, which removes the closed-loop decision feedback mechanism, and a CUC variant with only unidirectional prediction feedback. Table~\ref{tab:ablation} summarizes the results under different workload scales.

\textbf{Light Workload (140 tasks)}: 
When sufficient resources are available, CL-DFL achieves a violation rate of 0.00$\%$ 
while maintaining a resource utilization of resource utilization to 41.82$\%$. Compared with the ablation variants, CL-DFL improves utilization by approximately
6$\%$, demonstrating the effectiveness of neural-guided scheduling in reducing resource fragmentation.

\textbf{Saturated Workload (220 tasks)}: 
Under increased resource contention, the advantage of closed-loop adaptation becomes more evident. DPTO-Naive suffers from a violation rate of 6.58$\%$, exceeding the predefined SLA threshold, whereas CL-DFL reduces the violation rate to 
3.92$\%$ by adaptively adjusting confidence parameters through the DFL feedback mechanism. Meanwhile, CL-DFL maintains 100.00$\%$ user satisfaction and achieves a resource utilization of 56.54$\%$, indicating a better trade-off among violation rate, user satisfaction, and resource utilization.  

\textbf{High-Pressure Workload (300 tasks)}: Under highly congested workloads, all approaches experience increased violations due to limited resource availability. Nevertheless, CL-DFL achieves a lower violation rate of 28.58\% and improves resource utilization to 73.13$\%$, outperforming DPTO-Naive by 6.75$\%$. These results demonstrate that the integration of spatio-temporal perception, closed-loop decision-focused learning, and neural-guided scheduling jointly improves robustness under dynamic heterogeneous cloud workloads.

\begin{table}[h]
\centering
\caption{Performance Ablation Comparison Between CL-DFL and Baseline Variants Under Varying Task Scales (Azure 2023 Dataset).}
\label{tab:ablation}
\small
\begin{tabular}{l l c c c}
\toprule
Task Scale & Method & \multicolumn{1}{c}{$f_1$ (\%) $\downarrow$} & \multicolumn{1}{c}{$f_2$ (\%) $\uparrow$} & \multicolumn{1}{c}{$f_3$ (\%) $\uparrow$} \\
\midrule
\multirow{3}{*}{140 Tasks} & DPTO-Naive & \underline{0.58} & \textbf{100.00} & 35.81 \\
 & CUC & \textbf{0.00} & \textbf{100.00} & \underline{35.82} \\
 & CL-DFL & \textbf{0.00} & \underline{99.86} & \textbf{41.82} \\
\midrule
\multirow{3}{*}{220 Tasks} & DPTO-Naive & 6.58 & 99.97 & 52.91 \\
 & CUC & \textbf{2.08} & \underline{99.98} & \underline{54.76} \\
 & CL-DFL & \underline{3.92} & \textbf{100.00} & \textbf{56.54} \\
\midrule
\multirow{3}{*}{300 Tasks} & DPTO-Naive & 35.25 & \textbf{99.99} & 66.38 \\
 & CUC & \underline{31.25} & \textbf{99.99} & \underline{70.85} \\
 & CL-DFL & \textbf{28.58} & \underline{99.98} & \textbf{73.13} \\
\bottomrule
\end{tabular}\\
\begin{minipage}{0.9\linewidth}
\vspace{1mm}
\end{minipage}
\end{table}

\section{Conclusion}
This paper investigated the MOCOP of pre-collected task scheduling in heterogeneous cloud platforms and proposed the CL-DFL framework to address dynamic workloads and uncertain resource availability. By introducing a TPE-based zeroth-order DFL mechanism, CL-DFL establishes a closed-loop feedback pathway between prediction and scheduling, overcoming the limitations of traditional PTO paradigms. Moreover, the GNeuro-PLS strategy based on GRPO enables efficient Pareto solution searching in non-convex and black-box environments. Extensive experiments on real-world cloud workload traces demonstrate that CL-DFL achieves effective trade-offs among violation rate, user satisfaction, and resource utilization while maintaining low-latency decision-making. These results validate the effectiveness of CL-DFL for dynamic heterogeneous cloud resource orchestration. Future work will focus on extending the framework to distributed cross-domain computing environments and broader robust optimization scenarios.

\bibliographystyle{IEEEtran}
\bibliography{Reference}
\end{document}